\documentclass[9pt,journal]{IEEEtran}

\usepackage{bm}
\usepackage[inline]{enumitem}
\usepackage[minnames=2,maxnames=3,style=ieee,backend=bibtex]{biblatex}
\usepackage{hyperref}
\newcommand{\figref}[2][{}]{\hyperref[#2]{\figurename~\ref{#2}#1}}

\ifCLASSINFOpdf
   \usepackage[pdftex]{graphicx}
   \graphicspath{{./pdf/}{./jpeg/}}
   \DeclareGraphicsExtensions{.pdf,.jpeg,.png}
\else
   \usepackage[dvips]{graphicx}
   \graphicspath{{../eps/}}
   \DeclareGraphicsExtensions{.eps}
\fi

\usepackage{amsmath}

\usepackage{cleveref}
\usepackage{booktabs}
\usepackage{longtable}

\usepackage{array}

\ifCLASSOPTIONcompsoc
 \usepackage[caption=false,font=normalsize,labelfont=sf,textfont=sf]{subfig}
\else
 \usepackage[caption=false,font=footnotesize]{subfig}
\fi

\usepackage{stfloats}

\ifCLASSOPTIONcaptionsoff
 \usepackage[nomarkers]{endfloat}
\let\MYoriglatexcaption\caption
\renewcommand{\caption}[2][\relax]{\MYoriglatexcaption[#2]{#2}}
\fi

\usepackage{listings}
\usepackage{courier}

\usepackage{url}

\begin{document}
\title{Impact of Internal Algebraic Variable Treatment on Transient Stability Simulation Performance}

\author{Hantao~Cui,~\IEEEmembership{Senior~Member,~IEEE}%
}

\markboth{}%
{Cui, \MakeLowercase{\textit{et al.}}: Impacts of Algebraic Variable Treatment}

\maketitle

\begin{abstract}

It is a general notion that, in transient stability simulations, reducing the
number of algebraic variables for the differential-algebraic equations (DAE) can
improve the simulation performance. Many simulation programs split algebraic
variables internal to a dynamic model from the full DAE and evaluate them
outside each iterative step, using results from the previous iteration. The
updated internal variables are then treated as constants when solving for the
current iteration. This letter discusses how such a split formulation can impact
simulation performance. Case studies using various systems with synchronous
generator and converter models demonstrate the impact of the split on the
convergence pattern and simulation performance.
\end{abstract}

\IEEEpeerreviewmaketitle

\section{Introduction}
\IEEEPARstart{N}{umerical} simulation is a widely used yet computationally
challenging technique for power system transient stability assessment. The
simulation process essentially solves differential-algebraic equations (DAE) for
the network and dynamic devices. Algebraic variables are the instantaneous
quantities in the time horizon of electromechanical transient, and the
corresponding algebraic equations are hard constraints that the solutions need
to satisfy. Since common DAE solvers utilize Newton's method to solve the DAE or
the algebraic equations (AE), the number of algebraic variables can affect the
size of the AE and thus the computational performance.

The most well-known algebraic variables are the bus voltage phasors that
correspond to network equations. Also, there exist algebraic variables in
dynamic models (termed as ``internal algebraic variables") to describe physical
or mathematical relationships. In a generator, for instance, the bus voltage
projected to the \textit{d}-axis is
\begin{equation}
\label{eq:daxis}
v_d = v \cos(\delta - \theta)
\end{equation}
where $v$ is the voltage magnitude, $\theta$ is the bus phase angle, and
$\delta$ is the rotor angle. The algebraic variable $v_d$ and the equation is
trivial because one can substitute $v_d$ for the full equation to eliminate it.

Not all internal algebraic variables can be eliminated, because not all
algebraic variables have an explicit solution. For example, the stator
electrical equations for the two-axis generator is given by
\begin{equation}
\begin{array}{ll}
0 = v_q + r_a I_q - e'_q + x'_d I_d \\
0 = v_d + r_a I_d - e'_d - x'_q I_q
\end{array}
\end{equation}
where $e'_q$ and $e'_d$ are differential states for the rotor transient
voltages. One will not be able to eliminate $I_q$ and $I_d$ because they are
used to compute $\dot e'_q$ and $\dot e'_d$. In this example, $I_q$ and $I_d$
are algebraic variables internal to the generator, as opposed to bus voltages
that are shared across devices. For generality, the vector of internal algebraic
equations $\textbf{g}_i$ is given by
\begin{equation}
    \bm{0} = \bm{g}_i(\bm{x}, \bm{y}_i, \bm{y}_e)
\end{equation}
where $\bm{x}$ is the vector of state variables, and $\bm{y}_i$ and
$\bm{y}_e$ are the vectors of internal and external algebraic variables,
respectively.

There are two ways of treating internal algebraic variables and the
corresponding equations, resulting in two categories of DAE formulations. The
first category extends network equations to form the generalized algebraic
equations that include the internal ones \cite{milanoPowerSystemModelling2010}.
Assuming the simultaneous solution method, this formulation will incorporate the
derivative information of the internal variables in the full Jacobian matrix
and can thus obtain consistent solutions. This is termed the full DAE
formulation, given by

\begin{equation}
    \label{eq:fullDAE}
    \begin{array}{ll}
    \dot{\bm{x}} = \bm{f}(\bm{x}, \bm{y}_i, \bm{y}_e) \\
    \bm{0} = \bm{g}(\bm{x}, \bm{y}_i, \bm{y}_e)
    \end{array}
\end{equation}
where $\bm{g}$ is the compact notation for $\bm{g}_i$ and $\bm{g}_e$.

The second approach is to split internal algebraic equations from the full DAE
by rewriting internal algebraic variables in explicit forms. The split equations
will be evaluated using solutions from the previous iteration. Introduce a
superscript $k-1$ to denote the previous iteration, the internal variables are
given by
\begin{equation}
    \label{eq:splitDAE-internal}
    \bm{y}_i^{(k)} = \hat{\bm{g}}(\bm{x}_i^{(k-1)}, \bm{y}_i^{(k-1)}, \bm{y}_e^{(k-1)})
\end{equation}
which will be used to solve the following DAE:
\begin{equation}
    \label{eq:splitDAE-rest}
    \begin{array}{ll}
    \dot{\bm{x}}^{(k)} &= \bm{f}(\bm{x}^{(k)}, \bm{y}^{(k)}_i, \bm{y}^{(k)}_e) \\
    \bm{0} &= \bm{g}_e(\bm{x}^{(k)}, \bm{y}^{(k)}_i, \bm{y}^{(k)}_e)
    \end{array}
\end{equation}

Note that the solutions $\bm{x}^{(k)}, \bm{y}^{(k)}_i, \bm{y}^{(k)}_e$ from
\eqref{eq:splitDAE-internal}- \eqref{eq:splitDAE-rest} will not simultaneously
satisfy \eqref{eq:fullDAE}. In other words, there is a gap between the solutions
of \eqref{eq:fullDAE} and \eqref{eq:splitDAE-internal}-\eqref{eq:splitDAE-rest}.
The size of the gap depend on the variable scale and function characteristics, such
as linearity.

If all internal algebraic equations are split, then $\bm{g}_e$ becomes the
network equations, namely, $\bm{0} = \bm{I} - \bm{YV}$.
The split formulation is not unique. One can split all the internal
algebraic equations as in \cite{sauerPowerSystemDynamics2017} or split a subset,
which will be discussed in \Cref{sec:case-studies}.

This letter investigates the impacts of the two approaches to handling internal
algebraic variables on the simulation performance. \Cref{sec:characteristics}
briefly analyzes the characteristics of the two approaches.
\Cref{sec:case-studies} presents extensive case studies on representative small, medium, and
large systems using the two formulations on synchronous and renewable generators.
\Cref{sec:conclusions} concludes the study.

\section{Characteristics of the Full and the Split Formulations}
\label{sec:characteristics}

The split DAE formulation is widely used in commercial and open-source tools
\autocite{sauerPowerSystemDynamics2017} because of the following advantages:

\begin{enumerate}
    \item Simple to implement: only the explicit equation needs to be
    implemented. No partial derivatives are required, meaning that less
    programming is needed for the derivative functions.
    \item The same network equations and the derivatives can be used for the
    algebraic equations. It avoids the efforts to resize, compute, and assemble
    new Jacobian matrices.
\end{enumerate}

The full DAE is used in
\autocite{cuiHybridSymbolicNumericFramework2021c,milanoPowerSystemModelling2010}
with these characteristics:

\begin{enumerate}
    \item Consistent solutions can be obtained for states, and internal and
    external algebraic variables because the derivative information for all
    equations are reflected in the full Jacobian.
    \item The Jacobian matrix is larger due to the larger number of
    algebraic variables in the DAE. It leads to longer matrix factorization time
    since LU methods have a complexity of $\mathcal{O}(n^3)$.
\end{enumerate}

In terms of performance, there is a trade-off between a smaller matrix size for
quick factorization and fewer iterations for fewer function calls. Also, the
implementation complexity and the reuse of the network solver need to be considered.
Since there is no analytical formula to quantify such a trade-off, the impacts
of the two formulations will be studied by numerical simulations of systems of
various sizes.

\section{Case Studies}
\label{sec:case-studies}

Case studies are performed by splitting variables internal to the round-rotor
synchronous generator model (GENROU) and the generic renewable models (REGC\_A
and REEC\_A). Simulations are performed in the opensource ANDES tool
\autocite{cuiHybridSymbolicNumericFramework2021c} on Intel i9-10920X running
Debian 12. All test cases are available in the repository
\autocite{cuiANDESPythonSoftware2022}.

The implicit trapezoidal method is used to integrate the DAE, using a
convergence tolerance of $10^{-4}$. To reduce the number of matrix
factorizations, a ``dishonest" method is applied to rebuild and factorize the
Jacobian matrix every three iterations beyond the third. Also, to improve
convergence, the Jacobian matrix will be rebuilt and factorized honestly within
0.1 sec of disturbances.

\subsection{GENROU Model}

The split is performed on the flux linkage equations of

\begin{equation}
\begin{array}{lll}
    \psi_{aq} &=& \gamma_{q1} e'_{d}  + e''_{q} \left(1 - \gamma_{q1}\right) \\
    \psi_{ad} &=& \gamma_{d1} e'_{q} + \gamma_{d2} e''_{d} \left(x'_{d} - x_{l}\right) \\
    \psi_{a} &=& \sqrt{\psi_{ad}^{2} + \psi_{aq}^{2}}
\end{array}
\end{equation}
where $\gamma_{d1}$, $\gamma_{q1}$, $x'_d$, $x_l$ are parameters, and $e'_d$,
$e'_q$, $e''_d$ and $e''_q$ are state variables
\autocite{milanoPowerSystemModelling2010}. These equations are
necessary to compute $\psi_a$ as the input for the saturation function.

\subsubsection{IEEE 14-bus system} Five GENROU devices and multiple generator
controllers exist in this system. The full DAE model consists of 30 states and
119 algebraic variables, while the split model has 104 algebraic variables. The
simulated disturbances are a line trip at $t=0.1~\rm sec$ and a reconnection at
$t=0.2~\rm sec$.

\begin{figure}
    \centering
    \includegraphics{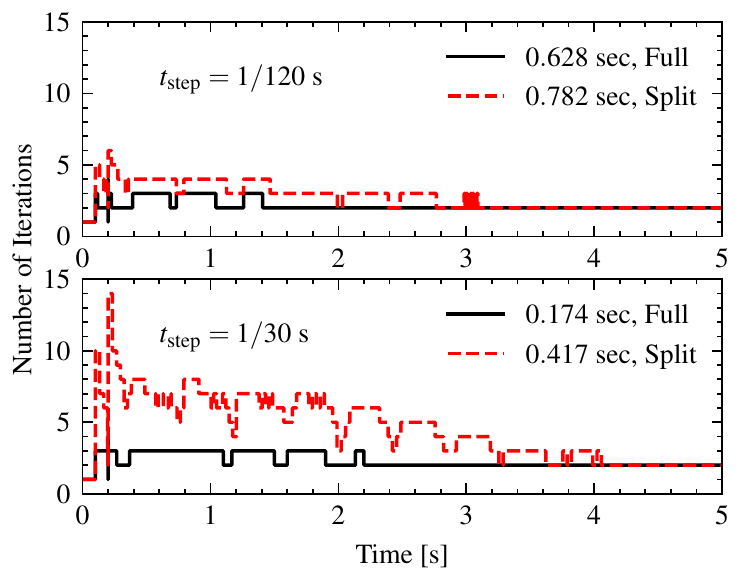}
    \caption{Statistics of the simulation performance for the IEEE 14-bus system using different integration step sizes.}
    \label{fig:ieee14-genrou-niter}
\end{figure}

\figref{fig:ieee14-genrou-niter} compares the number of iterations and
simulation time for the full and the  split formulations in the IEEE 14-bus
system. Two simulation step sizes are compared, namely, $1/120~\rm sec$ and $1/30~\rm
sec$, where the former is widely used in commercial tools. It is expected that
reducing the step size by a factor of four would result in four times the
computational load in residual building and equation solving but partially
compensated by the reduction in iteration number. Both the full and split
formulations with different step sizes yield the same dynamic response
trajectory and are thus omitted due to page limits.

The following characteristics are observed and analyzed:
\begin{enumerate}
    \item The more efficient approach is full DAE formulation with a relatively large
    step size. Even with 15 more algebraic variables, the full DAE is
    considerably and consistently faster.
    \item A small step size reduces the number of iterations due to the reduced
    gap between the split equations and the DAE. But it increases the
    computation time due to the step number increase.
    \item The split DAE benefits less than the full formulation from a larger
    step size due to the increased difficulty in convergence.
\end{enumerate}

\subsubsection{CURENT North America 944-bus model} This test system is a
medium-size case with a large variety of generator, exciter, turbine governor,
and power system stabilizer models. A total of 76 GENROU devices are in use. The
applied disturbances are a bus-to-ground fault at $t=0.1~\rm sec$ and its
clearance by a line trip at $t=0.2~\rm sec$.

\begin{figure}
    \centering
    \includegraphics{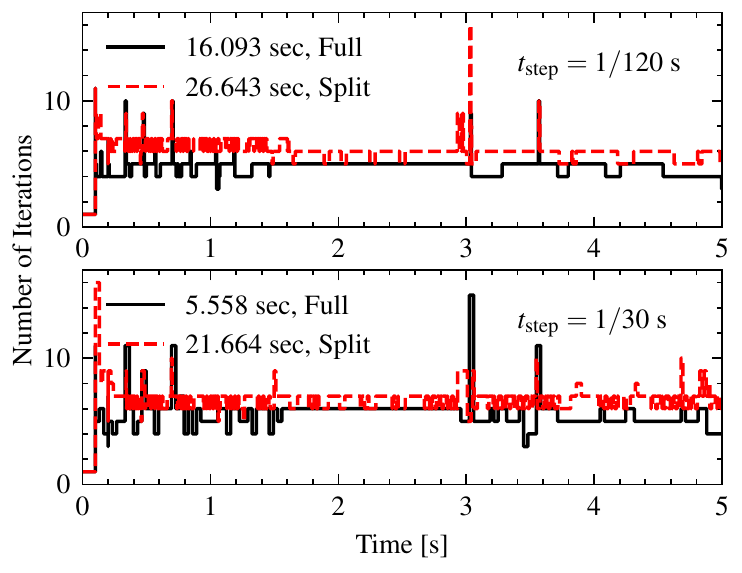}
    \caption{Simulation performance of the CURENT 944-bus system.}
    \label{fig:curent-na-niter}
\end{figure}

\figref{fig:curent-na-niter} shows the computational performance for the CURENT
system. Due to the model complexity such as limiters, the iteration counts do
not exhibit decreasing trend for the five seconds simulated. It can be observed
that the full DAE with a large step size is leading in performance. Also,
the split DAE benefits even less from the step size reduction, merely
reducing the time from $26.64~\rm sec$ to $  21.66~\rm sec$ due to the increased
number of iterations.

\subsubsection{Fictitious Polish 9241-bus system} This case is created from the
9241-bus system  \autocite{zimmermanMATPOWERSteadyStateOperations2011} by adding
generators, exciters, and turbine governors with generic parameters to represent
large systems. The full DAE of the system consists of 14,450 states and 61,833
algebraic variables, including those for 1,445 GENROU devices. The applied
disturbance is a generator trip at Bus 190 at $t=0.1~\rm sec$.

\begin{figure}
    \centering
    \includegraphics{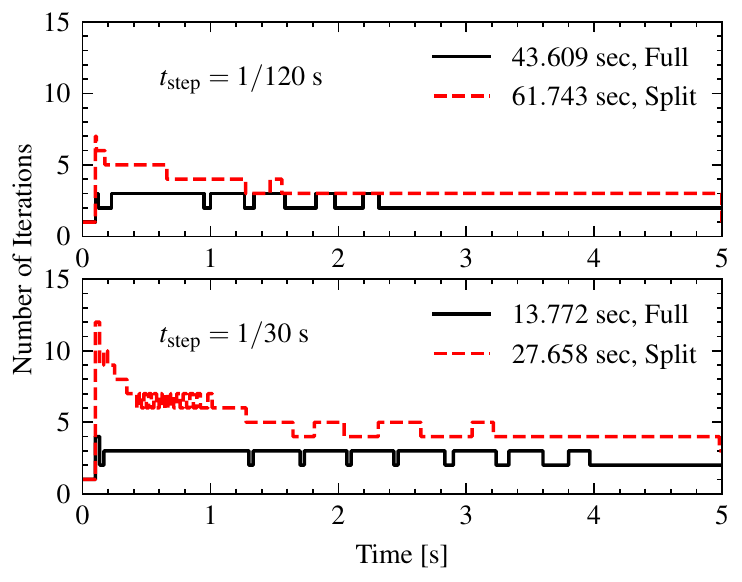}
    \caption{Simulation performance of the 9241-bus system.}
    \label{fig:case9241-niter}
\end{figure}

The full and split formulations yield the same transient trajectories for the two step
sizes. The computational performance follows the same observations as the
previous two cases. Comparing all the three test cases, we observe and
generalize the following:

\begin{enumerate}
    \item The full DAE can consistently scale by roughly a factor of three,
    namely, increasing the step size from $1/120~\rm sec$ to $1/30~\rm sec$
    reduces the run time by 3x. The factor of three is a result of roughly a
    quarter of the computational load that is offset by a slightly higher
    iteration count.
    \item The split DAE is less deterministic in terms of computational
    scalability. The speed-up factor for a quadruple step size reduces the
    run time by approximately half at most, depending on the system dynamics.
\end{enumerate}

\subsection{Renewable Generation Models}

The full and split formulation of the renewable energy converter model (REGC\_A)
and its electric control model (REEC\_A) are compared. For the REGC\_A model,
the low-voltage active power management output equations are split. For the
REEC\_A model, the voltage deviation equation, and the additional current
injection equations are split. Note that all these equations are linear.
Simulations are first performed using the IEEE 14-bus system, where 80\% of the
capacity on Bus 3 is replaced with two converters. A bus-to-ground fault is
applied to Bus 6, followed by a line trip to clear the fault.

\begin{figure}
    \centering
    \includegraphics{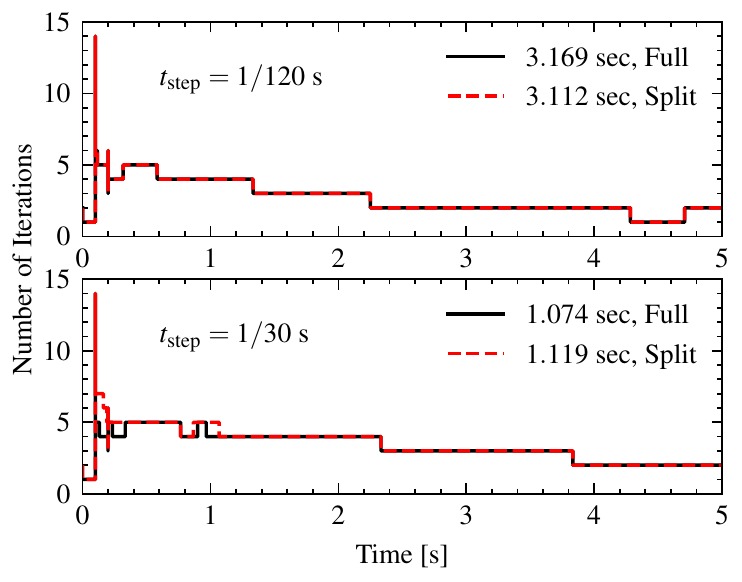}
    \caption{Simulation performance of the IEEE 14-bus system with renewables.}
    \label{fig:ieee14-regc-niter}
\end{figure}

\figref{fig:ieee14-regc-niter} plots the simulation performance for the two
formulations with different step sizes. In this case, the split formulation has
much less impact compared with those for the GENROU model. Still, for
$t=1/30~\rm sec$, multiple runs show that the split formulation is consistently slower by
several percent. For $t=1/120~\rm s$, the two formulations are close in performance
with no consistent winner.

The minor performance difference is due to the linearity of the equations that
are split, as well as the relatively small size of the test system. Input
changes to these equations result in linear corrections to the output, so that
the gap between the split variables and the full DAE solution can remain small.
Also, for $t=1/120~\rm sec$, given the same iteration count shown in
\figref{fig:ieee14-regc-niter}, the full DAE formulation with a larger DAE
require more time to build, factorize, and solve.

Next, the performance of the full and the split formulations are compared using
the 9241-bus system. A pair of renewable energy converters and electrical control
models are attached to each bus with a synchronous generator to substitute for
10\% of the active and reactive power outputs. The number of algebraic equations
for the full DAE and the split DAE is 116,743 and 110,963, respectively. Both
formulations have 28,900 differential states. A bus-to-ground fault is applied
at Bus 4 at $t=0.1~\rm sec$ and cleared at $t=0.2~\rm sec$.
\begin{figure}
    \centering
    \includegraphics{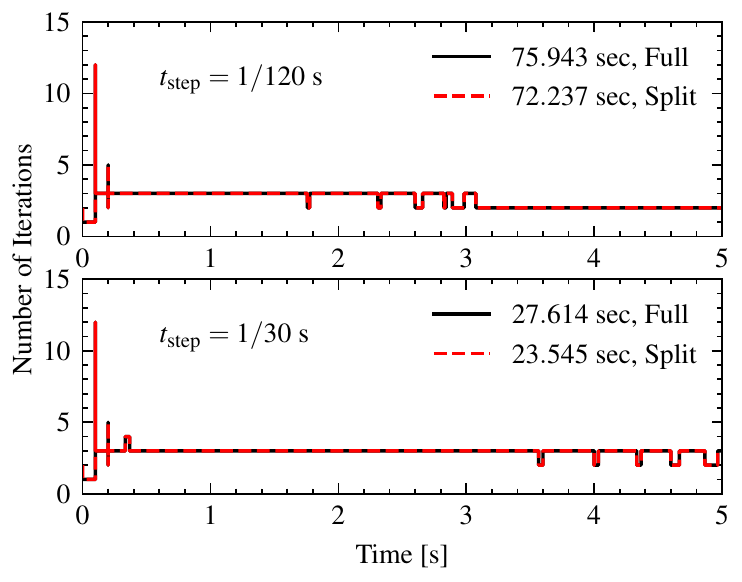}
    \caption{Simulation performance of the 9241-bus system with renewables.}
    \label{fig:case9241-regc-niter}
\end{figure}

\figref{fig:case9241-regc-niter} shows the performance results. In this case,
even for the step size of $1/30~\rm sec$, the convergence patterns are similar
for both formulations. In other words, the split of the linear equations does
not incur significant difficulty in convergence. Regardless of the step size,
the full DAE formulation is consistently slower than the split formulation due
to the larger size of the DAE.

\section{Conclusions}
\label{sec:conclusions}

This letter investigates the impact of the treatment of internal algebraic
variables on transient simulation performance. The impacts are studied on
synchronous generator models and renewable converter models in systems of various sizes.
Our conclusions are:

\begin{enumerate}
    \item The performance of the full and the split formulations
    depend on the equations being split and the size of the system.
    \item For nonlinear algebraic equations like the flux linkage equations the
    synchronous generators, the more computationally efficient approach is to keep
    the variables in the DAE so that the iteration count can remain low when a
    large step size is applied.
    \item Splitting linear equations, such as the ones in the converters, from
    the full DAE does not incur significant increases in the iteration count but can
    reduce the size of the Jacobian matrix, which is a significant factor for
    large systems.
\end{enumerate}

\ifCLASSOPTIONcaptionsoff
  \newpage
\fi

\printbibliography

\end{document}